\documentclass[runningheads]{llncs}
\usepackage[T1]{fontenc}
\usepackage{graphicx}
\usepackage{hyperref}
\usepackage{color}

\usepackage[labelsep=period, labelfont=bf]{caption}
\usepackage{subcaption}
\usepackage{booktabs}
\usepackage{multirow}
\usepackage{multicol}
\usepackage{array}
\usepackage{amsmath}
\usepackage[misc,geometry]{ifsym}
\usepackage{tablefootnote}
\usepackage{makecell}

\usepackage[most]{tcolorbox}
\definecolor{frame}{HTML}{ccefff}    
\definecolor{back}{HTML}{FAFEFF}     

\begin{document}
\title{Automated Assessment of Encouragement and Warmth in Classrooms Leveraging Multimodal Emotional Features and ChatGPT}
\titlerunning{Automated Assessment of Encouragement and Warmth in Classrooms}
%
\author{Ruikun Hou\inst{1,3}
    \and Tim Fütterer\inst{1}
    \and Babette Bühler\inst{1}
    \and Efe Bozkir\inst{1,3}
    \and Peter Gerjets\inst{2}
    \and Ulrich Trautwein\inst{1}
    \and Enkelejda Kasneci\inst{3}
}
\authorrunning{R. Hou et al.}
\institute{University of Tübingen, Geschwister-Scholl-Platz, 72074 Tübingen, Germany\\
    \email{\{ruikun.hou,tim.fuetterer,babette.buehler,efe.bozkir,\\
        ulrich.trautwein\}@uni-tuebingen.de}\\
    \and
    Leibniz-Instiut für Wissensmedien, Schleichstrasse 6, 72076 Tübingen, Germany\\
    \email{p.gerjets@iwm-tuebingen.de}\\
    \and
    Technical University of Munich, Arcisstrsse 21, 80333 Munich, Germany\\
    \email{\{ruikun.hou,efe.bozkir,enkelejda.kasneci\}@tum.de}
}
\maketitle              
\begin{abstract}
    Classroom observation protocols standardize the assessment of teaching effectiveness and facilitate comprehension of classroom interactions.
    Whereas these protocols offer teachers specific feedback on their teaching practices, the manual coding by human raters is resource-intensive and often unreliable.
    This has sparked interest in developing AI-driven, cost-effective methods for automating such holistic coding.
    Our work explores a multimodal approach to automatically estimating \textit{encouragement and warmth} in classrooms, a key component of the Global Teaching Insights (GTI) study’s observation protocol.
    To this end, we employed facial and speech emotion recognition with sentiment analysis to extract interpretable features from video, audio, and transcript data.
    The prediction task involved both classification and regression methods.
    Additionally, in light of recent large language models' remarkable text annotation capabilities, we evaluated ChatGPT's zero-shot performance on this scoring task based on transcripts.
    We demonstrated our approach on the GTI dataset, comprising 367 16-minute video segments from 92 authentic lesson recordings.
    The inferences of GPT-4 and the best-trained model yielded correlations of $r = .341$ and $r = .441$ with human ratings, respectively.
    Combining estimates from both models through averaging, an ensemble approach achieved a correlation of $r = .513$, comparable to human inter-rater reliability.
    Our model explanation analysis indicated that text sentiment features were the primary contributors to the trained model's decisions. Moreover, GPT-4 could deliver logical and concrete reasoning as potential teacher guidelines.
    Our findings provide insights into using advanced, multimodal techniques for automated classroom observation, aiming to foster teacher training through frequent and valuable feedback.

\end{abstract}
\section{Introduction}
A comprehensive understanding of classroom interactions is crucial to deciphering the quality of teaching, providing hence the opportunity to foster an educational environment where learning thrives \cite{seidel2007teaching}. This understanding paves the way for interventions like real-time feedback, enriching the teaching and learning processes and empowering researchers to dissect teaching scenarios with heightened reliability and efficiency. These insights are particularly pivotal when assessing facets of teaching effectiveness, such as student support, where elements like classroom encouragement and warmth are not mere niceties but essential catalysts for effective teaching \cite{pianta2009conceptualization}. The ability to delve deeper into these components promises to enhance educational practices and refine teacher professional development programs, steering them toward fostering these nurturing classroom atmospheres.
Traditionally, the task of capturing the nuances of teaching dynamics has involved human observers, employing structured classroom observation protocols like CLASS (Classroom Assessment Scoring System \cite{pianta2008classroom}).
For this task, human observers watch lesson recordings and assign scores based on teaching effectiveness measures defined in the protocols.
Whereas a human rating approach is valuable, it is fraught with multiple challenges \cite{demszky2023can}. Human-based observations are inherently subjective, often leading to higher-inference holistic-level assessments with low inter-rater agreement. Moreover, the manual nature of these assessments makes them resource-intensive in terms of time and cost.

Against this backdrop, there is growing interest in developing AI-driven approaches to automatically coding classroom observation protocols. Prior research addressed the task by either employing multimodal feature extraction together with supervised classifiers \cite{james2018inferring,ramakrishnan2021toward} or relying on advanced large language models (LLMs) \cite{wang2023chatgpt,whitehill2023automated}.
In line with existing studies, our goal is to make an initial contribution to automatic evaluation approaches that reflect the eye of a highly trained human evaluator but overcome the limitations of human subjectivity and resource constraints.
We focus on a specific aspect of teaching effectiveness, namely Encouragement and Warmth (EW), a significant component in the observation protocol of the Global Teaching Insights (GTI) study \cite{gti2020}. The component involves the provision of encouragement for students throughout their work, including positive comments and compliments, along with moments of shared warmth such as smiling and laughter \cite{gti2018trainingnotes}.
This corresponds to the Positive Climate (PC) dimension in CLASS.

To assess classroom EW, we investigate the use of both multimodal models tailored to domain-specific data and LLMs' generative capabilities, aiming to harness each method's unique strengths. We first propose a supervised-learning approach based on multimodal representations of emotion and then apply the Shapley additive explanations (SHAP) \cite{lundberg2017unified} technique to examine the contributions of these explicit features. Additionally, we explore whether recent LLMs like ChatGPT, relying on their zero-shot annotation capabilities, can effectively score EW based on classroom discourse and reasonably interpret their decisions. Lastly, we evaluate the predictive performance of an ensemble approach that combines estimates from supervised models and ChatGPT.

\section{Related Work}
Recently, the success of machine learning has triggered a growing trend towards AI applications in classroom settings, such as analysis of student behavior \cite{ahuja2019edusense,buhler2023automated} and engagement \cite{goldberg2021attentive,sumer2021multimodal}, classroom discourse \cite{hunkins2022beautiful,jensen2021deep}, as well as teacher perception \cite{sumer2018teachers}.
Specifically, a few recent studies have targeted the holistic analysis of automated teaching effectiveness coding within classroom observation protocols. They can be categorized into two strands: multimodal supervised methods \cite{james2018inferring,ramakrishnan2021toward} and LLM-based methods \cite{wang2023chatgpt,whitehill2023automated}.

The first multimodal machine-learning system was proposed by James et al. \cite{james2018inferring}, which employed visual, conversational, and acoustic features to identify whether the classroom climate is positive following the CLASS protocol. Their trained binary classifier yielded a $F1$-score of 0.77.
Furthermore, Ramakrishnan et al. \cite{ramakrishnan2021toward} presented a multimodal architecture to achieve a more fine-grained scoring of both PC and Negative Climate (NC) dimensions in CLASS. In line with the 7-point coding scale of CLASS, the prediction task was formulated as a 7-class classification problem.
They utilized an ensemble model integrating visual and auditory pathways, achieving correlations between predictions and human ratings of \textit{r} = .55 (PC) and \textit{r} = .63 (NC).

In addition to multimodal approaches, recent research investigated using LLMs for transcript-based classroom observation scoring.
Wang and Demszky \cite{wang2023chatgpt} pioneered the employment of ChatGPT's zero-shot capabilities to score classroom transcripts across various dimensions of teaching effectiveness.
They prompted GPT-3.5 with an entire transcript segment and a description of the respective dimension requiring rating. Based on 100 authentic transcript segments, ChatGPT estimates resulted in a weak correlation of \textit{r} = .04 with human-coded scores regarding CLASS PC.
Instead of using complete transcripts as prompts, Whitehill and LoCasale-Crouch \cite{whitehill2023automated} introduced an LLM-based approach focusing on utterance-level analysis. They leveraged zero-shot prompting with an LLM to distinguish individual teacher utterances and further trained a linear regressor on aggregated session-level features to assess the CLASS Instructional Support domain. Their best-performing model, employing features concatenated from the LLM and a Bag-of-Words method, reached a correlation of \textit{r} = .46. Although this utterance-level method approached human inter-rater reliability, it lacked in grasping the semantic context in dialogues.

We propose leveraging both multimodal models and LLMs to predict classroom observation scores.
Our work stands for a further contribution to this relatively unexplored domain, extending prior studies in the following aspects:
(1) We focus on extracting interpretative features that explicitly constitute EW-related behavioral indicators, as opposed to utilizing low-level auditory features \cite{james2018inferring,ramakrishnan2021toward}. This enables us to (2) apply explainability frameworks to understand which behaviors contribute to model predictions, which is central in practical applications, such as teacher training. (3) Additionally, we explore regression methods that account for the ordering attribute of data labels compared to standard classification.
Moreover, GPT-3.5 resulted in subpar performance for CLASS PC scoring \cite{wang2023chatgpt}. As the recent GPT-4 model has demonstrated improved text understanding and generation capabilities as well as reduced hallucination \cite{achiam2023gpt}, (4) we evaluate if GPT-4 surpasses its predecessor in achieving adequate zero-shot performance for this particular scoring task.
(5) Furthermore, we investigate the potential of an ensemble method to boost predictive accuracy by leveraging the strengths of supervised models and ChatGPT's zero-shot approaches.

\section{GTI Dataset}
\label{sec:dataset}

The dataset employed in this work stems from GTI \cite{gti2020}, a large-scale classroom video study aiming to achieve a profound understanding of teaching and learning worldwide. Across eight participating countries, the study centers on a shared pedagogical topic in mathematics (quadratic equations) and emphasizes objective evidence on classroom practice by directly observing authentic lesson videos and instructional materials.
A video observation protocol was developed to ensure consistent rating processes within the study. At a high level, the protocol consists of six \textit{domains}: Classroom Management, Social-Emotional Support, Discourse, Quality of Subject Matter, Student Cognitive Engagement, and Assessment of and Responses to Student Understanding. Each domain comprises multiple \textit{components} measuring the quality of distinct teaching constructs at higher inference levels.
Human raters observed instructional videos divided into 16-minute segments and assigned each segment a score on a 4-point scale for each component, guided by associated behavioral examples. The raters were required to attend dedicated training lessons and engage in several quality control checks, thus yielding heavy workloads to guarantee rating reliability.
We focus on the EW component from the Social-Emotional Support domain, which captures behavioral characteristics comparable to the CLASS PC dimension. Particularly, Encouragement refers to using positive verbal and nonverbal cues to inspire students to begin or persist in tasks, such as reassurance for students' mistakes, positive comments, and compliments on their work, while Warmth is represented by, e.g., smiling, laughter, joking, and playfulness \cite{gti2018trainingnotes}. The score scale from one to four aligns with the occurrence frequency of these behaviors from no evidence to frequent instances throughout one segment.

\begin{figure}[h]
    \centering
    \begin{subfigure}[t]{0.45\textwidth}
        \centering
        \includegraphics[width=0.55\textwidth]{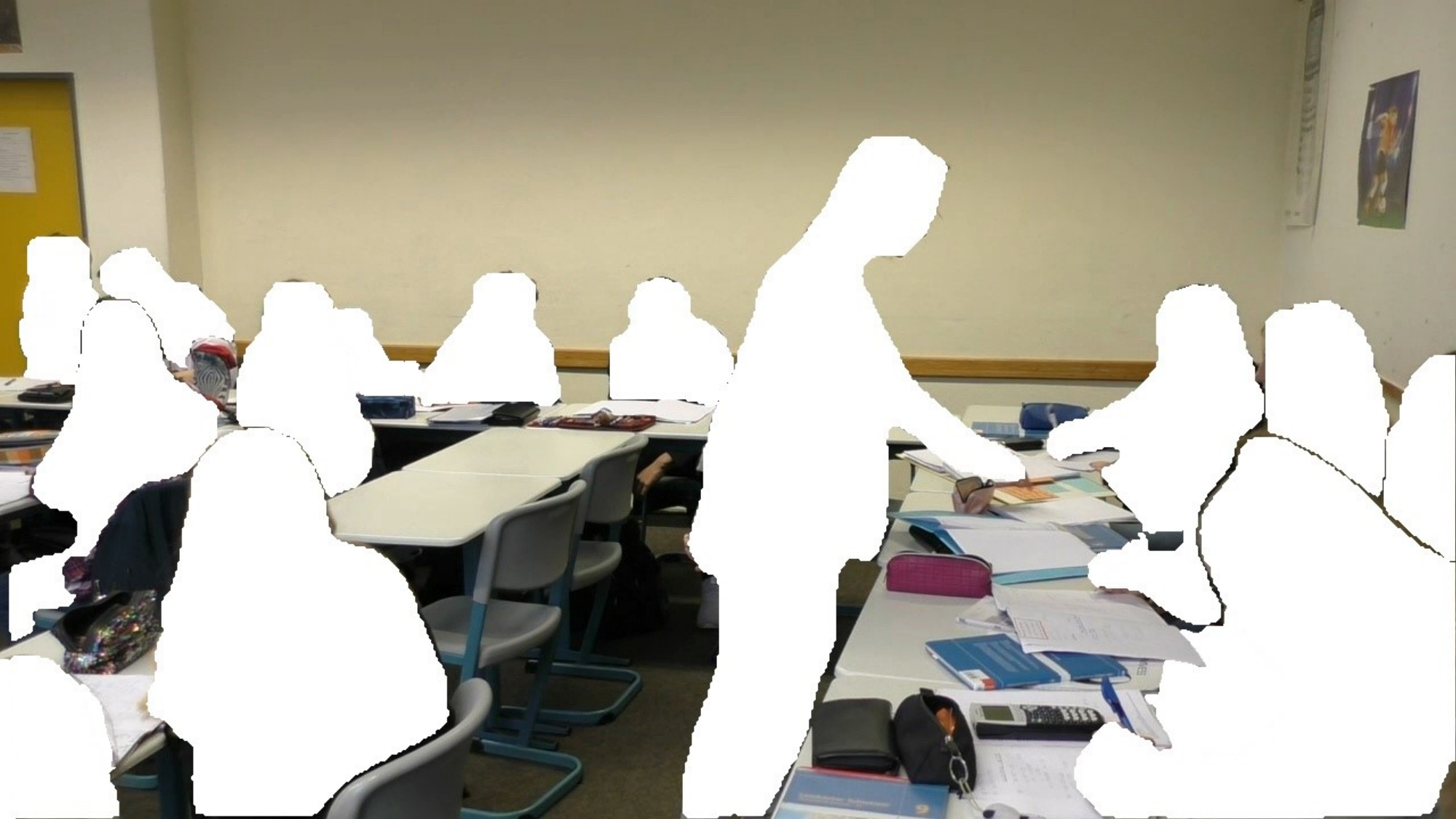}
        \caption{Teacher camera}
        \label{fig:classroom_sample_frame_teacher}
    \end{subfigure}
    \begin{subfigure}[t]{0.45\textwidth}
        \centering
        \includegraphics[width=0.55\textwidth]{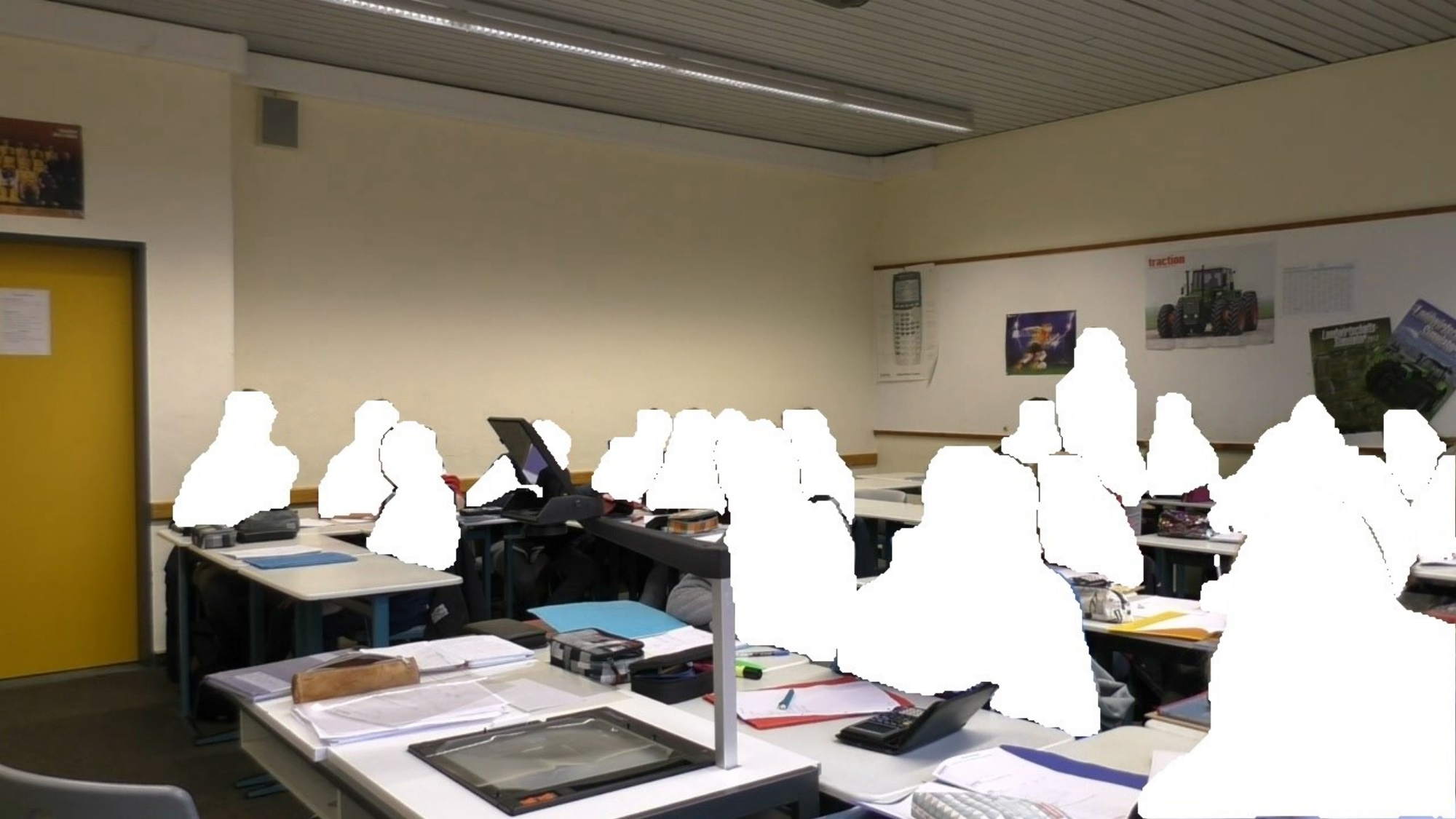}
        \caption{Student camera}
        \label{fig:classroom_sample_frame_student}
    \end{subfigure}
    \caption{Classroom frame from two cameras, with people erased for privacy.}
    \label{fig:classroom_sample_frame}
\end{figure}
We used the GTI data collected in Germany, involving 100 video-recorded math lessons, 50 recruited teachers, and over 1,140 students. Due to data protection regulations, we were only allowed to access 92 of the recordings. Each lesson lasted from 40 to 90 minutes. The lessons were videotaped simultaneously by two cameras at 25 FPS (frames per second): One tracked the teacher's movements (Fig.~\ref{fig:classroom_sample_frame_teacher}), while the other was stationary and positioned to capture as many students as possible (Fig.~\ref{fig:classroom_sample_frame_student}). We utilized the recordings from the latter camera, where the frontal faces of most participants were visible. In addition, GTI supplied lesson transcripts created by human transcribers, which we employed as the text modality in this work. In transcripts, timestamps and speakers were annotated following every conversation turn, where speakers were anonymized by their IDs, such as "L" for the teacher and "S01" for a student. The rating process involved 14 raters in Germany, with each lesson being annotated by two randomly assigned raters.
Following the GTI protocol, we preprocessed the data by splitting lesson recordings and transcripts into 16-minute segments. If the last segment of one recording spanned less than eight minutes, it was merged with the preceding segment. This resulted in 367 segments, serving as the dataset on which we built and evaluated our automated estimation methods.

\section{Methodology}

\begin{figure}[h]
    \centering
    \includegraphics[width=\textwidth]{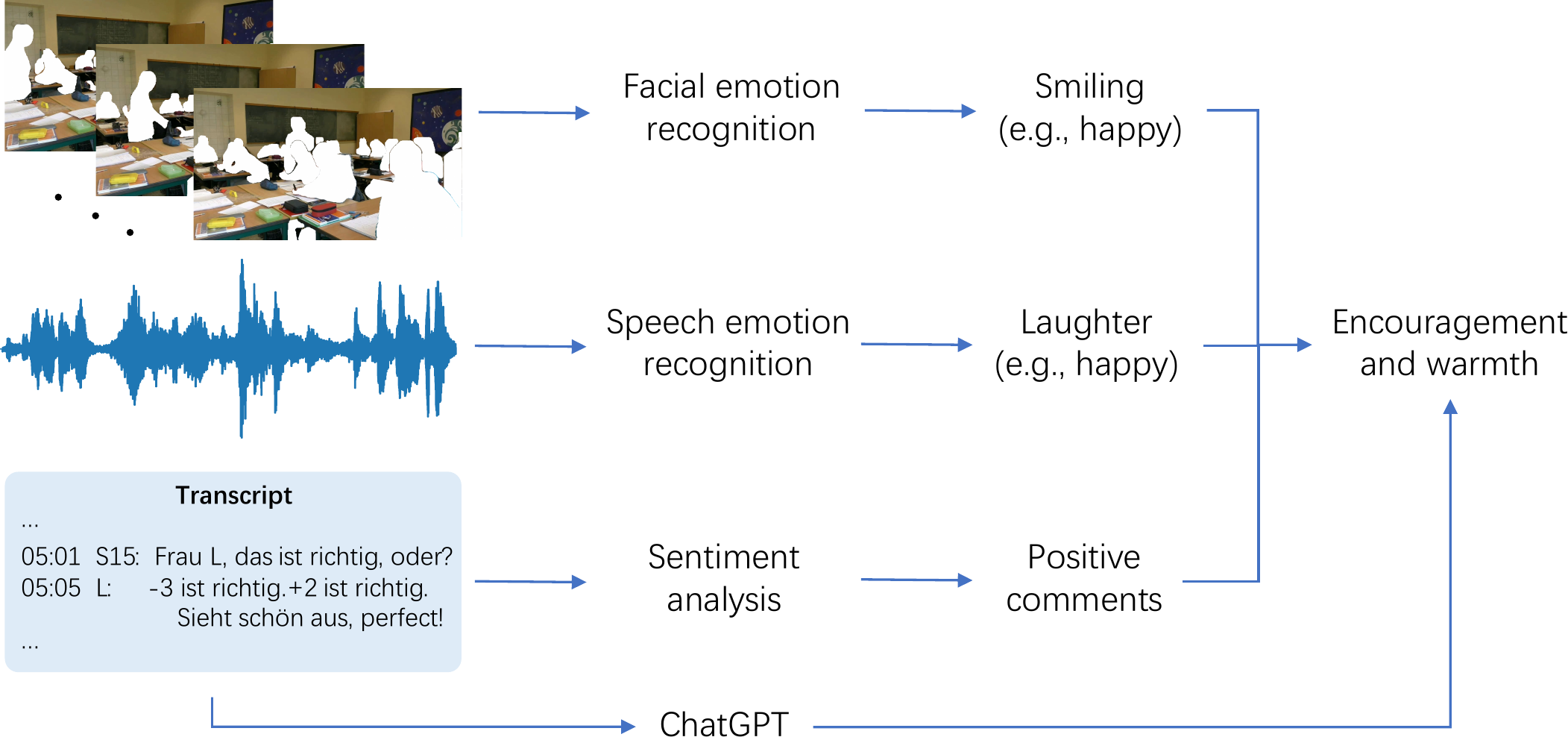}
    \caption{Pipeline for multimodal estimation of EW scores.}
    \label{fig:pipline}
\end{figure}

In this section, we elaborate on our proposed approach to automated estimation of EW scores (Fig.~\ref{fig:pipline}), including supervised learning methods based on multimodal features extraction (Sect.~\ref{subsec:multimodal}), ChatGPT zero-shot annotation (Sect.~\ref{subsec:chatgpt}), and an ensemble model combining both paths (Sect.~\ref{subsec:ensemble}).

\subsection{Multimodal Supervised Models}
\label{subsec:multimodal}
We aimed to build a machine-learning approach to mimic a human's rating process to the greatest extent possible. Thus, we focused on the GTI rater training materials (e.g., coding rubrics and guidelines), which played an essential role in helping raters comprehend rating specifications. As indicated in the training materials, raters were required to pay attention solely to the dedicated behaviors listed in the EW definition (see Sect.~\ref{sec:dataset}). Hence, we extracted interpretable features by employing off-the-shelf techniques to represent the EW-associated behaviors. Notably, these behavioral cues are typically linked to the affective states of teachers and students. Therefore, we implemented facial emotion recognition to identify "smiling" in videos, applied speech emotion recognition (SER) to detect "laughter" in audio, and carried out text sentiment analysis to distinguish "positive comments" in classroom discourse.

\subsubsection{Facial Emotion Recognition}
We adopted a deep neural network architecture, EmoNet \cite{toisoul2021estimation}, which performs multi-task predictions of facial emotion in a single pass. Beyond the recognition of discrete emotion categories, EmoNet simultaneously estimates continuous valence (positive or negative) and arousal (excited or calm) values defined in the circumplex model of affect \cite{russell1980circumplex}, enabling a more comprehensive depiction of human emotions.
We utilized an EmoNet model pre-trained on AffectNet \cite{mollahosseini2017affectnet}, a dataset containing a vast amount of facial images in the wild annotated with discrete and continuous emotion labels.
To apply the model to our classroom dataset, we first down-sampled video segments from 25 to 2 FPS (i.e., 1920 frames for a 16-minute segment), which considered the minimal variation between consecutive frames and reduced computational resources. We then employed RetinaFace \cite{deng2020retinaface}, known for its competitive performance in crowded environments, to detect faces in each frame. Afterward, each face crop was input into EmoNet, which predicted valence and arousal values ranging from -1 to 1, along with a probability distribution over five emotions (neutral, happy, sad, surprise, fear). For each frame, we aggregated the predictions by averaging the valence and arousal values as well as the estimation scores over five discrete labels across all detected faces, yielding a 7D feature vector.

\subsubsection{Speech Emotion Recognition}
Recent work \cite{pepino21interspeech} tackled SER by employing transfer learning based on embeddings derived from pre-trained deep models, showing superior performance over conventional methods that relied on low-level acoustic features.
Given that the spoken language in our dataset is German, we utilized XLSR proposed by Facebook AI \cite{conneau2020unsupervised} to extract cross-lingual deep embeddings from raw audio signals. XLSR is pre-trained in over 50 languages and built on top of wav2vec 2.0 \cite{baevski2020wav2vec}, a transformer-based model trained on unlabeled data in a self-supervised manner for speech recognition. This approach enhances performance for low-resource languages.
In particular, we used a publicly released XLSR model fine-tuned in German, mapping a speech signal to a 1024D latent embedding. We applied the model as a feature extractor to EmoDB \cite{burkhardt2005database}, a database consisting of 535 audio instances, each representing a German utterance categorized into seven discrete emotion labels (anger, boredom, disgust, fear, happiness, sadness, neutral). Based on the resulting features, we trained a two-layer Multi-Layer Perceptron (MLP) classifier on the randomly selected 80\% of the EmoDB data and used the remainder for testing. The accuracy on the test set achieved 0.95.
In our classroom setup, we divided the 16-minute audio segments into 192 5-second windows without overlap due to the observation that over 95\% of the EmoDB instances lasted under 5 seconds, with an average duration of 2.8 seconds. We then applied the XLSR model and the trained MLP classifier to infer emotions expressed in each audio window, yielding a 7D feature vector as a probability distribution over the seven emotion labels.

\subsubsection{Sentiment Analysis}
TextBlob\footnote{https://textblob.readthedocs.io/} is a well-established toolkit supporting multiple natural language processing functionalities, such as tokenization and translation. It conducts lexicon-based sentiment analysis utilizing a predefined polarity dictionary. Prior research employed the toolkit for sentiment analysis of student feedback \cite{sadriu2021automated}.
In this work, we applied TextBlob-de\footnote{https://textblob-de.readthedocs.io/}, a German language extension, to the transcript segments that were manually annotated regarding turn-taking between speakers. The tool assessed the sentiment of each teacher or student utterance by assigning a polarity score ranging from -1 to 1. In line with \cite{sadriu2021automated}, we categorized utterances according to their polarity scores, labeling those above/equal to/below 0 as positive/neutral/negative, respectively. We generated a 4D feature vector for each transcript segment by counting positive, neutral, and negative utterances and computing a cumulative polarity score across all utterances. Since each lesson recording's final segment typically did not span 16 minutes, we normalized the feature vector by the respective segment duration.

\subsubsection{Prediction}
It is noteworthy that both facial and speech emotion features were temporal, whereas text sentiment features were generated in a segment-wise way. We aggregated the visual and auditory features for each segment by computing the mean along the temporal dimension. We then concatenated the segment-wise features from all three modalities into a single 18D vector.
Afterward, we formulated the EW estimation problem as both classification and regression tasks. To compare the two approaches fairly, we trained an identical set of models: Random Forest (RF), Support Vector Machine (SVM), and MLP with two layers.
As each segment was double-rated, we calculated the average of the two human ratings as the ground truth. For classification, we rounded the ground truth in the training set to the nearest integer, within the range of one to four.
To guarantee generalization, all models were evaluated through stratified, lesson-independent 5-fold cross-validation, such that segments from the same lesson were always grouped into the same fold, and each fold maintained the original score distribution. The evaluation results were averaged across all test folds. Before model fitting, we standardized the features using the mean and the standard deviation computed from the training data. Moreover, we carried out grid-search hyperparameter tuning to identify the best-performing configuration for each model.

\subsection{ChatGPT Zero-Shot Annotation}
\label{subsec:chatgpt}
\begin{figure}[]
    \centering
    \includegraphics[width=1\textwidth]{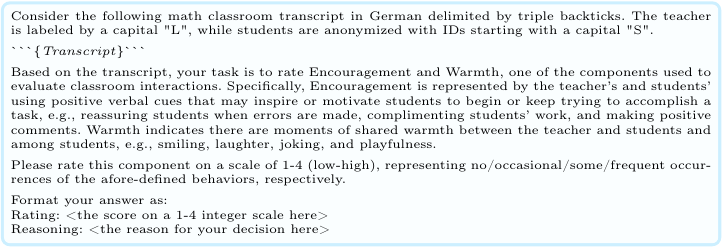}
    \caption{Prompt for ChatGPT.}
    \label{fig:gpt_prompt}
\end{figure}

Recent advances in LLMs introduce vast opportunities for their application in the education sector to enhance teaching and learning \cite{kasneci2023chatgpt,sessler2023peer}.
LLMs are typically trained on extensive text corpora and thus equipped with broad knowledge, enabling quick adaptation to new tasks without the need for retraining. Such zero-shot capability has shown notable effectiveness across various text annotation problems \cite{gilardi2023chatgpt}.
In the GTI study, transcripts play a significant role in rating processes (e.g., raters in many countries employed shorthand and highlighting tools directly on transcripts). Therefore, we explore the potential of ChatGPT to assess the EW component without requiring training, relying exclusively on classroom transcripts. In particular, we prompted ChatGPT with a transcript segment to rate, along with EW's definition, behavioral examples, and coding rubrics, as depicted in Fig.~\ref{fig:gpt_prompt}.
Besides, we requested ChatGPT to reason its decision on the score assignment, as suggested by \cite{wang2023chatgpt}. For this task, we employed two GPT models via the OpenAI API to compare their performance, namely \textit{gpt-3.5-turbo-1106} and \textit{gpt-4-1106-vision-preview}.
To consistently compare with those trained models, we evaluated ChatGPT's zero-shot performance on the same five test folds and averaged the results.

\subsection{Ensemble Model}
\label{subsec:ensemble}
For text feature extraction, the sentiment analysis was performed at the level of individual utterances.
ChatGPT analyzed an entire transcript, thus leveraging contextual information for a comprehensive understanding of classroom discourse \cite{whitehill2023automated}. Considering that each method might provide distinct and potentially complementary insights, we constructed an ensemble model that integrates the trained model with the ChatGPT zero-shot approach. The ensemble model computes the weighted average of estimates from both base models. Weights are allocated according to their evaluation performance on the training set in each fold. The unweighted averaging was also tested for comparison.

\section{Results}
\label{sec:results}

\subsection{Model Performance}
Similar to a recent study \cite{whitehill2023automated}, we treat automatic estimation approaches as individual raters and explore the extent of their consistency with human raters in terms of Pearson correlation coefficient $r$.
For this purpose, we first computed human inter-rater reliability (IRR) in a leave-one-rater-out fashion. In our context, there were 14 human raters conducting double ratings. We computed $r$ for each rater by comparing their ratings with those assigned by other raters. The coefficients obtained were then averaged across all raters. Meanwhile, we report standard error estimates (standard deviation over all raters divided by $\sqrt{14}$).
For model evaluation, we calculated the average of $r$ between model predictions and human ratings over five test folds, as well as the corresponding standard error estimates.
The results are summarized in Table~\ref{tab:results}. Except for GPT-3.5, the correlations for each model in each fold are statistically significant ($p <$ .05).

\begin{table}[t]
    \centering
    \caption{Results of GTI Encouragement and Warmth estimation.}
    \label{tab:results}
    \begin{tabular}{p{3.9cm}p{5.5cm}p{2.2cm}}
        \toprule
        \textbf{Approach}                      & \textbf{Model}                & \textbf{Pearson $r$}   \\
        \toprule
        Inter-Rater Reliability                & Human Raters                  & \textbf{0.513 (0.028)} \\
        \midrule
        \multirow{3}{*}{Multimodal Classifier} & RF                            & 0.391 (0.041)          \\
                                               & SVM                           & 0.375 (0.058)          \\
                                               & MLP                           & \textbf{0.392 (0.040)} \\
        \midrule
        \multirow{3}{*}{Multimodal Regressor}  & RF                            & 0.429 (0.051)          \\
                                               & SVM                           & 0.433 (0.041)          \\
                                               & MLP                           & \textbf{0.441 (0.039)} \\
        \midrule
        \multirow{2}{*}{ChatGPT Zero-Shot}     & GPT-3.5                       & 0.027 (0.071)          \\
                                               & GPT-4                         & \textbf{0.341 (0.037)} \\
        \midrule
        \multirow{2}{*}{Ensemble}              & MLP Reg.$+$GPT-4 (Unweighted) & 0.499 (0.033)          \\
                                               & MLP Reg.$+$GPT-4 (Weighted)   & \textbf{0.513 (0.036)} \\
        \bottomrule
        \multicolumn{3}{p{\textwidth}}{\textit{Notes.} For humans, $r$ corresponds to the average across raters. For automatic models, $r$ is the average over 5 folds between estimates and mean human ratings. Standard error estimates are shown in parentheses, with the best model for each approach in bold.}
    \end{tabular}
\end{table}

For multimodal supervised approaches, the best-trained model is the MLP regressor ($r$ = .441). Overall, it seems advantageous to formulate the problem as a regression task. This is evident from the fact that classifiers achieve the highest Pearson $r$ of .392, trailing behind all regressors in performance.
Another interesting result is that the two-layer MLP outperforms the other two conventional models in both classification and regression tasks, underscoring the efficacy of contemporary neural networks. It is noteworthy that adding more layers did not enhance the performance, potentially attributed to the risk of overfitting inherent in deeper models when confronted with a limited dataset size.
When turning to ChatGPT zero-shot results, a clear difference between the two model generations is identified. The estimates of GPT-3.5 show no significant correlation with human-rated scores. In contrast, GPT-4 has remarkably enhanced the capability to understand text, particularly achieving superior zero-shot performance over its predecessor to score EW in classroom discourse. It attains a Pearson $r$ of .341 without requiring any training.
Finally, the ensemble model, combining the best-performing MLP regressor with GPT-4, boosts the performance significantly. Using a simple averaging method raises the Pearson $r$ to .499. When the averaging process considers the respective reliability of both base models, the correlation reaches a peak of .513, identical to that of the human IRR.

\begin{figure}[t]
    \centering
    \includegraphics[width=.75\textwidth]{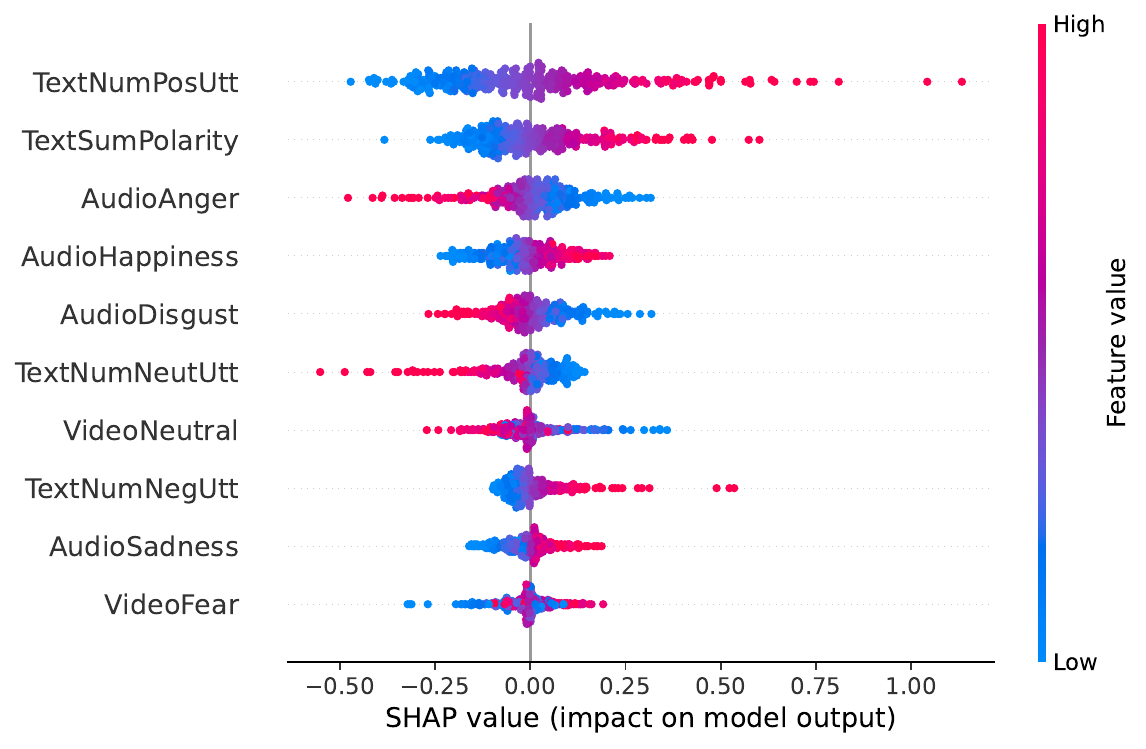}
    \caption{SHAP summary plot for MLP regressor. Points depict SHAP values per feature per data sample. Features are ranked by their importance (sum of SHAP value magnitudes over all samples). The 10 most influential features are shown.}
    \label{fig:shap}
\end{figure}

\subsection{Model Explanation}
\subsubsection{Supervised Models}
To explore which explicit features from which modality are influential in model decisions, we further applied Shapley additive explanation (SHAP) \cite{lundberg2017unified} to our MLP regressor. A SHAP value quantifies the impact of a feature on an individual prediction.
In the cross-validation setting, we gathered SHAP values from each test fold and then created a summary plot involving the entire dataset.
As illustrated in Fig.~\ref{fig:shap}, the group of verbal and auditory features appears to be more informative than those derived from videos in enabling the regressor to predict EW scores. Specifically, the feature describing the number of positive utterances within a transcript segment contributes the most, followed by the overall polarity and three speech emotion features. By analyzing the relationship between feature values and SHAP values illustrated in the plot, it is apparent that high values (shown in red) in the two most important features are associated with positive SHAP values, suggesting that more positive sentiment cues contribute to increasing the predicted EW score.
Additionally, the plot reveals the negative impact of detected anger and disgust as well as the positive impact of detected happiness in the audio on the prediction of higher EW scores.

\subsubsection{ChatGPT}
GPT-4's efficacy is evident not only in its agreement with human raters but also in its capability to provide logical reasoning. For example, GPT-4 assigned a score of 4 to a transcript consistently with the human ratings and explained its decision by identifying relevant evidence from the discourse:
\begin{quote}\scriptsize
    "[...] For instance, the teacher praises S15’s work as `sieht schön aus perfekt'\textit{(`looks beautiful perfect')} and encourages S04 by validating their thinking process. The teacher’s tone is patient and nurturing, especially visible in exchanges like `keine Panik'\textit{(`no panic')} [...] The teacher often uses humor, as seen in statements like `bevor hier einer weint'\textit{(`before someone cries here')} [...]"
\end{quote}
This aligns with the human rating procedure that measures the frequency of dedicated behaviors. Conversely, we observed that GPT-3.5 typically presented broad-level reasoning lacking concrete examples, which limited its explainability.

\section{Discussion}
Human raters achieve moderate agreement ($r =$ .513) for EW coding in our dataset. Prior work \cite{ramakrishnan2021toward} reported Pearson $r$ of .38 and .42 for CLASS PC in two datasets they utilized. This indicates that coding teaching effectiveness requires holistic analysis and higher-level inference.
Compared to human subjectivity, automatic tools benefit from providing more objective insights at scale \cite{demszky2023can,jensen2021deep}. To this end, we explore a novel approach to automated EW assessment.
Our methods achieve correlations of .441 (the best-trained model) and .341 (GPT-4 zero-shot annotation) between estimates and human ratings.
Combining both methods yields the highest predictive accuracy ($r =$ .513) on par with human IRR.
The findings from supervised models show that a set of multimodal, explicit, and low-dimensional features can effectively capture EW-relevant signals in a 16-minute lesson segment.
Unlike classifiers that treat categories independently, regressors account for the ordering attribute inherent in the data labels.
For example, mistaking a true score of one as four incurs a higher penalty than confusing it with two. This may explain the observed superior performance of regressors. Another reason is that the ground-truth labels were rounded to integers used as classification categories, leading to a loss of information.
Additionally, the SHAP analysis highlights the importance of text features, which can be interpreted by the EW coding rubrics where verbal cues constitute a large proportion of the associated behavioral indicators. Meanwhile, auditory features contribute more than visual features, which aligns with prior research findings \cite{ramakrishnan2021toward}.
Besides the competitive accuracy achieved by supervised models, GPT-4's ability to deliver persuasive reasoning showcases its potential as an easily accessible tool for teachers to obtain valuable feedback on classroom climate.
Further, the efficacy of the ensemble approach suggests that the two base models may capture complementary information. Integrating GPT-4 zero-shot annotations with a specialized and shallow model does not require resource-intensive fine-tuning of LLMs yet enhances the final performance, providing insights into strategies for using LLMs both effectively and efficiently when addressing similar tasks.

Due to the use of distinct datasets and protocols, we note that the results are not comparable to prior studies \cite{james2018inferring,ramakrishnan2021toward,wang2023chatgpt,whitehill2023automated} which focused on CLASS dimensions. GTI EW exhibits similar behavioral indicators to CLASS PC, but their coding rubrics differ in aspects such as score scale (4- vs. 7-point). Our exploration contributes to a broader understanding of similar constructs across different protocols.
Further, we explore methods for German speech and text processing, diverging from the common use of English data in existing research, which enriches the discussion on developing educational technologies in multilingual contexts.
Moreover, transcripts not only provide more informative features for predictive models but also serve as a privacy-conscious modality compared to video and audio recordings.
As classroom recordings are sensitive data involving minors, ethical considerations are of utmost importance. We emphasize that automated observation tools intend to streamline post hoc analysis and reduce the need for manual coding instead of being used for real-time classroom monitoring.

Our work is subject to several limitations.
First, our analysis utilized manual transcripts. We could implement advanced speech recognition techniques, toward a fully automatic system.
In classroom environments, students' faces far away from the camera lead to limited spatial resolution, reducing the performance of emotion recognition models. Solutions could be applying super-resolution methods to improve face image quality.
Due to the potential disagreement between distinct model explanation methods \cite{swamy2022evaluating}, it is valuable to validate various explainers on our EW scoring task.
Another approach to feature importance analysis would be an ablation study to compare the performance using unimodality.
Furthermore, one future direction is to adapt the proposed methods for the automated coding of additional GTI aspects such as classroom discourse to achieve a more comprehensive measurement of teaching effectiveness.
As GPT-4 is a multimodal LLM capable of image understanding, another future study involves exploring whether the inclusion of classroom frames together with transcripts could enhance estimation accuracy.
To further improve model generalizability, conducting cross-dataset evaluation, particularly employing data from diverse countries or cultures, would be part of future research.

\section{Conclusions}
We explore a machine-learning approach that harnesses multimodal supervised models and LLMs' zero-shot capabilities to automatically assess Encouragement and Warmth in classrooms.
The results indicate that our approach achieves rating performance on par with human inter-rater reliability.
We further show that verbal features contribute the most to model predictions, and GPT-4 provides specific evidence for its scoring decisions, outperforming GPT-3.5.
Such AI-driven methods have the potential to replicate and augment human observational capabilities in the future, enabling frequent and valuable feedback for educational researchers regarding several teaching effectiveness facets.

%
%
\bibliographystyle{splncs04}
\bibliography{mybibliography}

\end{document}